\documentclass[journal,twoside,web]{ieeecolor}
\usepackage{tmi}
\usepackage{cite}
\usepackage{amsmath,amssymb,amsfonts}
\usepackage{algorithmic}
\usepackage{graphicx}
\usepackage{textcomp}
\usepackage{booktabs}
\usepackage{multirow}
\usepackage{orcidlink}

\def\BibTeX{{\rm B\kern-.05em{\sc i\kern-.025em b}\kern-.08em
    T\kern-.1667em\lower.7ex\hbox{E}\kern-.125emX}}
\begin{document}
\title{APRF: Anti-Aliasing Projection Representation Field for Inverse Problem in Imaging}
\author{Zixuan Chen,~%\IEEEmembership{Member,~IEEE}
    % Huajun Zhou,~%\IEEEmembership{Member,~IEEE}
        Lingxiao Yang,~\IEEEmembership{Member,~IEEE},~
        Jianhuang Lai\textsuperscript{\orcidlink{0000-0003-3883-2024}},~\IEEEmembership{Senior Member,~IEEE},\\
        and Xiaohua Xie\textsuperscript{\orcidlink{0000-0002-0310-4679}}~\IEEEmembership{Member,~IEEE}% <-this % stops a space
\thanks{Manuscript received XXX XX, XXXX; revised XXX XX, XXXX. This work was supported in part by the National Natural Science Foundation of China under Grant 62072482.
}
\thanks{(Corresponding author: Xiaohua Xie.) All the authors are with the School of Computer Science and Engineering, Sun Yat-sen University, Guangzhou 510006, China; and with the Guangdong Province Key Laboratory of Information Security Technology, Guangzhou 510006, China; and also with the Key Laboratory of Machine Intelligence and Advanced Computing, Ministry of Education, Guangzhou 510006, China. (e-mail: chenzx3@mail2.sysu.edu.cn; yanglx9@mail.sysu.edu.cn; stsljh@mail. sysu.edu.cn; xiexiaoh6@mail.sysu.edu.cn.)}% <-this % stops a space
}

\maketitle

\newcommand{\bt}[1]{\textbf{#1}}
\newcommand{\df}[1]{\footnotesize{\textcolor{subsectioncolor}{#1}}}
\newcommand{\ua}{$\uparrow$}
\newcommand{\da}{$\downarrow$}
\newcommand{\ck}{$\checkmark$}

\begin{abstract}
Sparse-view Computed Tomography (SVCT) reconstruction is an ill-posed inverse problem in imaging that aims to acquire high-quality CT images based on sparsely-sampled measurements.
Recent works use Implicit Neural Representations (INRs) to build the coordinate-based mapping between sinograms and CT images.
However, these methods have not considered the correlation between adjacent projection views, resulting in aliasing artifacts on SV sinograms.
To address this issue, we propose a self-supervised SVCT reconstruction method -- Anti-Aliasing Projection Representation Field (APRF), which can build the continuous representation between adjacent projection views via the spatial constraints.
Specifically, \textit{APRF} only needs SV sinograms for training, which first employs a \textit{line-segment sampling} module to estimate the distribution of projection views in a local region, and then synthesizes the corresponding sinogram values using a \textit{center-based line integral} module.
After training \textit{APRF} on a single SV sinogram itself, it can synthesize the corresponding dense-view (DV) sinogram with consistent continuity.
High-quality CT images can be obtained by applying re-projection techniques on the predicted DV sinograms.
Extensive experiments on CT images demonstrate that \textit{APRF} outperforms state-of-the-art methods, yielding more accurate details and fewer artifacts.
Our code will be publicly available soon.
\end{abstract}

\begin{IEEEkeywords}
Imaging Inverse Problem, Sparse-View Computed Tomography (SVCT), Implicit Neural Representation (INR), Self-Supervised Learning, Anti-Aliasing.
\end{IEEEkeywords}
\section{Introduction}
\IEEEPARstart{C}{omputed} Tomography (CT) is a diagnostic imaging procedure that uses a combination of X-rays to observe the internal structure of the scanned objects.
It consists of two steps: \textit{i)} emitting X-rays in a circle around the scanned subjects and storing the information about attenuation properties at each projection angle in the sinograms; \textit{ii)} using re-projection techniques like \cite{FBP,ART} to transform the sinograms into CT images.
Acquiring high-quality CT images requires densely-sampled projection views, which means that subjects must be scanned for long periods of time without moving.
Exposure to such prolonged ionizing radiation may increase the risk of cancer in subjects \cite{CT}.
Consequently, Sparse-View Computed Tomography (SVCT) reconstruction, \emph{i.e.,} reconstructing CT images based on sparsely-sampled measurements, can significantly reduce the ionizing radiation from CT scans, which is of great concern in the field of public healthcare.

\begin{figure}[!t]
    \centering
    \includegraphics[width=0.48\textwidth]{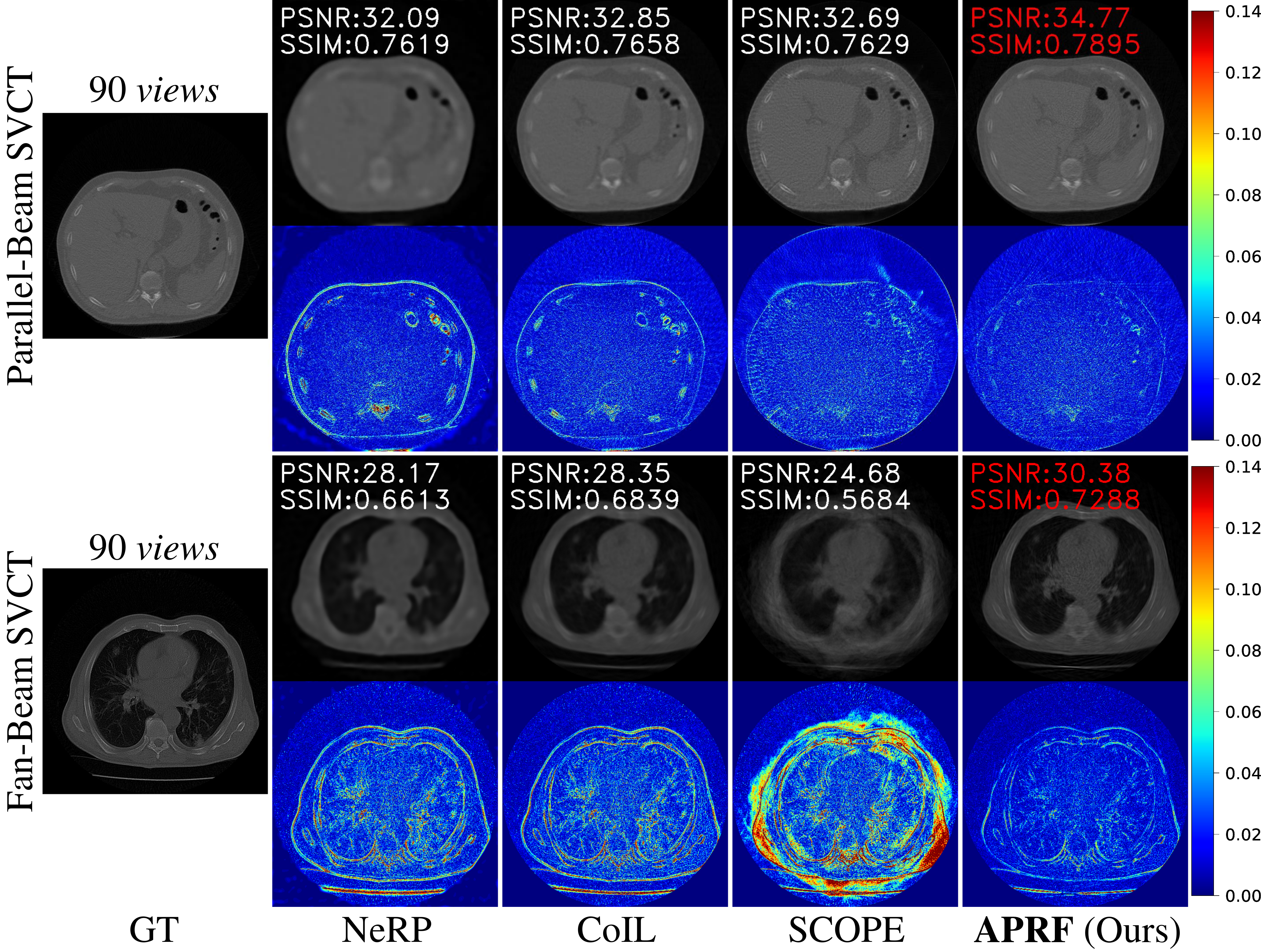}
    \caption{
      The visual examples of \textbf{2D} SVCT reconstruction results between our \textit{APRF} and the state-of-the-art INR-based methods: NeRP \cite{NeRP}, CoIL \cite{Coil} and SCOPE \cite{scope} on COVID-19 \cite{covid19} dataset.
      Heatmaps at the bottom visualize the difference related to the GT images.
      Red text indicates the best performance.
    }
    \label{fig:examples}
\end{figure}

\begin{figure*}[!t]
    \centering
    \includegraphics[width=\textwidth]{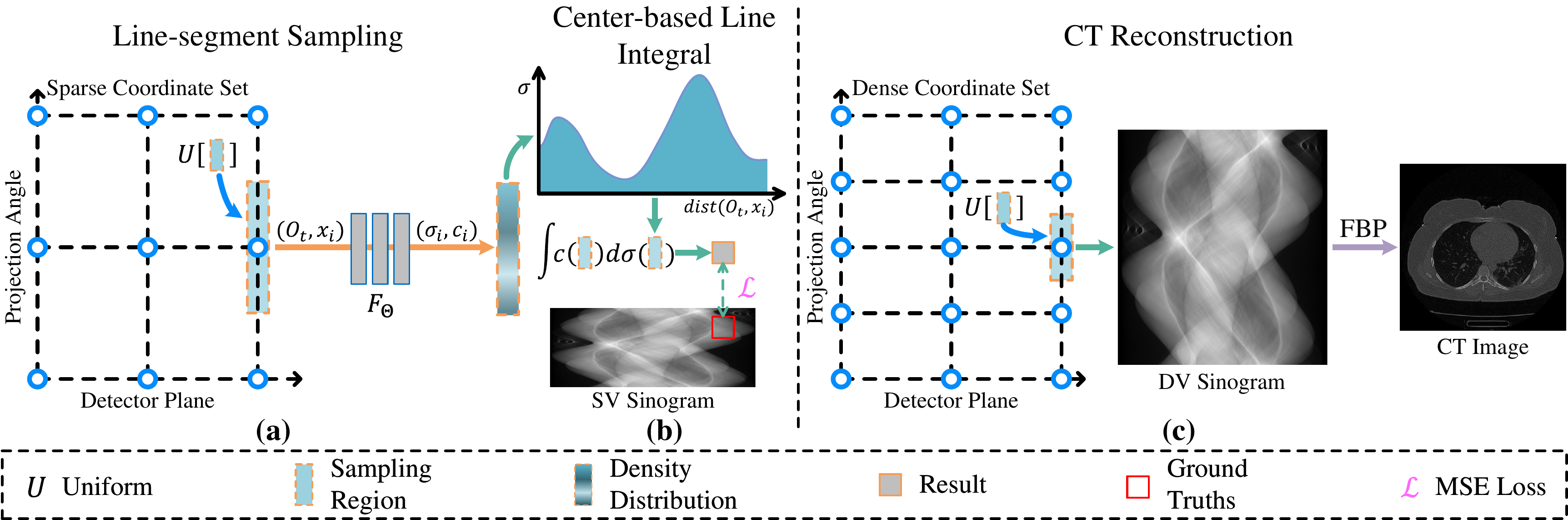}
    \caption{
      The overall architecture of the proposed \textit{APRF}.
      In the training stage, \bt{(a)} \textit{APRF} first uniformly samples $N$ points within the line-segment regions $\ell(O_{t}, \rho)$.
      Each point coordinate $x_{i}\in\{x_i\}^N_{i=1}$ and the line-segment center $O_t$ are fed into an MLP $F_{\Theta}$ to produce the corresponding density $\sigma_i\!=\!\sigma(O_t,x_i)$ and intensity $c_i\!=\!c(x_i)$.
      \bt{(b)} Assuming the distribution of the line-segment spaces is only related to the distance $dist(O_t, x_i)$, \textit{APRF} employs a line integral to predict the sinogram results.
      Since the proposed \textit{center-based line integral} is differentiable, \textit{APRF} can be optimized by the MSE loss in Eq. \ref{eq:loss} between the predicted results and SV sinogram pixels.
      In the test stage, \bt{(c)} \textit{APRF} first synthesizes DV sinograms and then applies FBP \cite{FBP} on the predicted results to acquire the target CT images.
    }
    \label{fig:overall}
\end{figure*}

SVCT reconstruction is a highly ill-posed inverse problem.
Since directly using the conventional re-projection techniques \cite{FBP} and \cite{ART} may produce severe artifacts, early studies \cite{art4svct,art4svct2} combined \cite{ART} with the regularization terms for optimization.
Subsequently, with the development of deep learning, many methods \cite{FBPConvNet,TFUnet,tmi2,tmi3,tmi4,tmi6,tmi1,tmi5} proposed different Convolutional Neural Networks (CNNs) to learn the mapping between SV sinograms and CT images.
These CNN-based methods are fully-supervised, requiring considerable SV sinogram and CT image pairs for training.
Recently, implicit neural representation (INR) techniques \cite{NeRF,LIIF} have yielded impressive performance in computer vision for numerous tasks.
Many studies \cite{Coil,scope,GRFF,NeRP} utilize the INR techniques to handle the SVCT reconstruction challenges.
One representative strategy is to build the coordinate-based field that directly represents the sinograms \cite{Coil}.
The other strategy is to build the coordinate-based representation field of CT density, and then synthesize the corresponding sinograms based on the differentiable \textit{Radon transformation} \cite{scope,GRFF,NeRP}.
However, these INR-based methods have not considered the spatial correlations between adjacent projections, producing blurry contents \cite{NeRP} and severe artifacts \cite{Coil,scope} (see Figure \ref{fig:examples}).

To address the above issues, we propose an Anti-Aliasing Projection Representation Field (APRF), which aims to model the correlation between adjacent projection views from the SV sinograms.
Unlike the above CNN-based methods, our \textit{APRF} is a self-supervised SVCT reconstruction method, which is trained on the SV sinograms without the reference of CT images.
Specifically, we propose a \textit{line-segment sampling} module to randomly sample $N$ projection angles within a spatial region.
To synthesize corresponding sinogram values, we also propose a \textit{center-based line integral} module, which employs a numerical integral on the sampled coordinates.
After optimization, the internal regions between adjacent projection views can be indirectly modeled by the spatial constraints, and thus \textit{APRF} can yield the dense-view (DV) sinogram with consistent continuity.
By applying re-projection techniques like \cite{FBP} on the predicted DV sinograms, our \textit{APRF} can acquire high-quality CT images (see Figure \ref{fig:examples}).
Comprehensive experiments on COVID-19 \cite{covid19} and KiTS19 \cite{kits19} datasets demonstrate the superiority of our \textit{APRF}, whose reconstruction quality surpasses state-of-the-art methods on \textit{parallel-}, \textit{fan-} and \textit{cone-}beam SVCT images.

In summary, the main contributions are:

\begin{itemize}
    \item We discover that existing INR-based methods are vulnerable to aliasing errors in the projection domains, and propose an anti-aliasing SVCT reconstruction method.  
    \item We propose a \textit{line-segment sampling} module and a \textit{center-based line integral} module to alleviate aliasing errors via spatial constraints.
    \item We conduct a series of experiments to demonstrate the effectiveness of our model.
\end{itemize}

\section{Related Works}

In this section, we first briefly review the Implicit Neural Representation (INR) techniques and their derived applications in computer vision and medical image reconstruction.
We then introduce some impressive advances in sparse-view computed tomography (SVCT) reconstruction, including some very recent INR-based methods.

\subsection{Implicit Neural Representation}
Modeling a continuous representation from discretely sampled data is a long-standing problem in image reconstruction.
Recent studies propose implicit neural representation (INR) techniques to solve this ill-posed problem, which aim to build a coordinate-based representation field from the collected samples.
Specifically, these INR-based methods use the Multi-Layer Perceptron (MLP) to encode the coordinates and learn the mapping between coordinates and training samples.
After training, based on the image continuity priors, these coordinate-based representation fields can produce the \textit{i.i.d} samples corresponding to the training set.
INR techniques have yielded impressive advances in image reconstruction for numerous tasks: single image super-resolution \cite{LIIF}, video super-resolution \cite{videoinr}, novel view synthesis \cite{NeRF}, generative modeling \cite{graf,Giraffe,dreamfusion}, and editing  \cite{nerf-editing,sine}.
Recently, some researchers have tried to adapt INR techniques to the medical domain.
Corona-Figueroa \emph{et al.} \cite{mednerf} adapted \cite{graf} to synthesize novel projection views with training on multi-view projection images, while Chen \emph{et al.} \cite{cunerf} proposed the cube-based modeling strategy to reform NeRF \cite{NeRF} for upsampling medical images at arbitrary scales.

\subsection{Sparse-View Computed Tomography Reconstruction}
Sparse-View Computed Tomography (SVCT) reconstruction aims to acquire CT images based on sparsely-sampled measurements, which can significantly reduce ionizing radiation.
Initial studies proposed analytical reconstruction methods: the Filtered Back-Projection (FBP) \cite{FBP}, and the iterative methods: \cite{ART} to transform the sinograms into CT images.
However, these methods have difficulty in obtaining high-quality CT images from the sparse-view (SV) sinograms and produce severe artifacts in their results.
To eliminate the artifacts, \cite{art4svct,art4svct2} combined \cite{ART} with the regularization terms for optimization.
With the advent of Convolutional Neural Networks (CNNs), \cite{FBPConvNet,TFUnet,tmi2,tmi3,tmi4,tmi6,tmi1,tmi5} built the CNN-based models to address the challenges of SVCT reconstruction.
\cite{FBPConvNet,TFUnet,tmi2,tmi3} reformed the U-Net \cite{U-net} architecture to learn the mapping between the sparse- and full-view CT pairs on a large dataset, while \cite{tmi4,tmi6} designed dense blocks to fuse hierarchical features.
Recent works \cite{Coil,GRFF,intratomo,NeRP,scope} use INR techniques to construct the coordinate-based mapping between SV sinograms and CT images.
These INR-based methods can be divided into two groups: \textit{i)} \cite{GRFF,intratomo,NeRP} simulate the density field of CT images and synthesize SV sinograms via differentiable FBP \cite{FBP} techniques for optimization; \textit{ii)} \cite{Coil,scope} build the coordinate-based representations of SV sinograms and apply FBP \cite{FBP} on the synthesized DV sinograms to acquire CT images.
% However, the above INR-based methods have some limitations. 
% A major drawback is that they have not modeled the correlation between adjacent projection views.
% Therefore, they suffer from aliasing errors in the projection domains, resulting in blurred results \cite{NeRP} and severe artifacts \cite{Coil,scope} (see Figure \ref{fig:examples}).

\section{Method}
In this section, we first analyze the reasons why existing INR-based methods suffer from aliasing errors in the projection domains.
Subsequently, we propose Anti-Aliasing Projection Representation Field (APRF) to address this problem, which consists of two modules: \textit{line-segment sampling} and \textit{center-based line integral}.
These two modules are introduced in the following subsections, and we also demonstrate why the aliasing errors can be eliminated by them.
Finally, we report the details of our methods, including spatial normalization and the architecture of Multi-Layer Perceptron (MLP).

The overall architecture is depicted in Figure \ref{fig:overall}.
As shown, in the training stage, \textit{APRF} trains the models on a single SV sinogram itself.
\textit{APRF} employs the proposed \textit{line-segment sampling} module \bt{(a)} and \textit{center-based line integral} module \textbf{(b)} to learn the mapping between the sparsely-sampled coordinates and the corresponding sparse-view (SV) sinogram pixels.
In the test stage, given a dense coordinate set, \textit{APRF} first synthesize the corresponding DV sinograms via \bt{(a)} and \bt{(b)}, and then applies FBP \cite{FBP} on the predicted results to reconstruct CT images.

\subsection{Analysis on Aliasing Errors in the Projection Domains}
\begin{figure}[!t]
    \centering
    \includegraphics[width=0.48\textwidth]{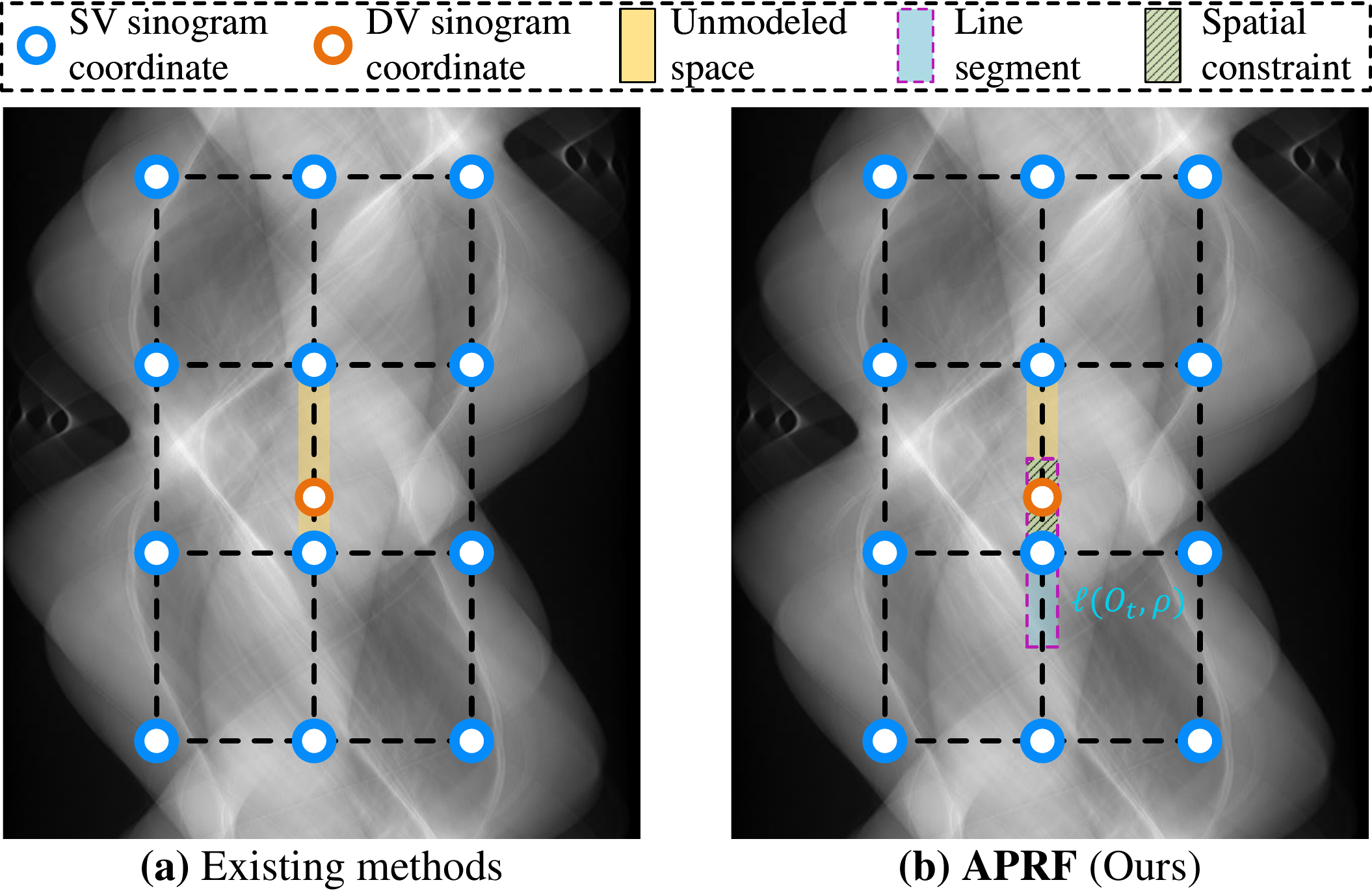}
    \caption{
        \bt{(a)} Existing INR-based methods only build the mapping between the coordinates and SV sinogram pixels.
        However, since they have not modeled the correlation between adjacent projections, they may leave some unmodeled spaces within these internal regions, leading to aliasing errors in the projection domains.
        \bt{(b)} By contrast, \textit{APRF} samples a line segment $\ell(O_{t}, \rho)$ instead of a single point to synthesize the corresponding SV sinogram pixel, which can build the continuous representation between adjacent projection views via the spatial constraints during optimization.
    }
    \label{fig:analysis}
\end{figure}

Existing INR-based SVCT reconstruction methods aim to build the mapping between position coordinates and sinogram pixels.
However, we discover that these methods have not considered building the correlation between adjacent projections, leading to aliasing errors in the projection domains.
We provide a vivid example in Figure \ref{fig:analysis} \bt{(a)} to explain our findings.
As shown, these INR-based methods can only model the points corresponding to sparsely-sampled projection angles, while the internal regions between adjacent projection views have never been directly modeled during training.
These unmodeled spaces may cause aliasing errors in building dense projection views, yielding blurred results \cite{GRFF,NeRP} and severe artifacts \cite{Coil,scope} (see Figure \ref{fig:examples}).

To alleviate aliasing errors, we argue that it is required to build the correlation between adjacent projection views.
As shown in Figure \ref{fig:analysis} \bt{(b)}, sampling a line-segment region instead of a single point can build the distance-based correlation within the unmodeled spaces. 
After optimization, the model can build the continuous representation between adjacent projections, reducing the aliasing errors.

To achieve our motivation, we propose \textit{line-segment sampling} module to sample projection angles within a line-segment region.
We also propose a \textit{center-based line integral} module to synthesize corresponding sinogram values.
After optimization,  the internal regions between adjacent SV projection views can be indirectly modeled by the spatial constraints, and thus \textit{APRF} can yield the dense-view (DV) sinogram with consistent continuity.

\begin{figure*}[!t]
    \centering
    \includegraphics[width=\textwidth]{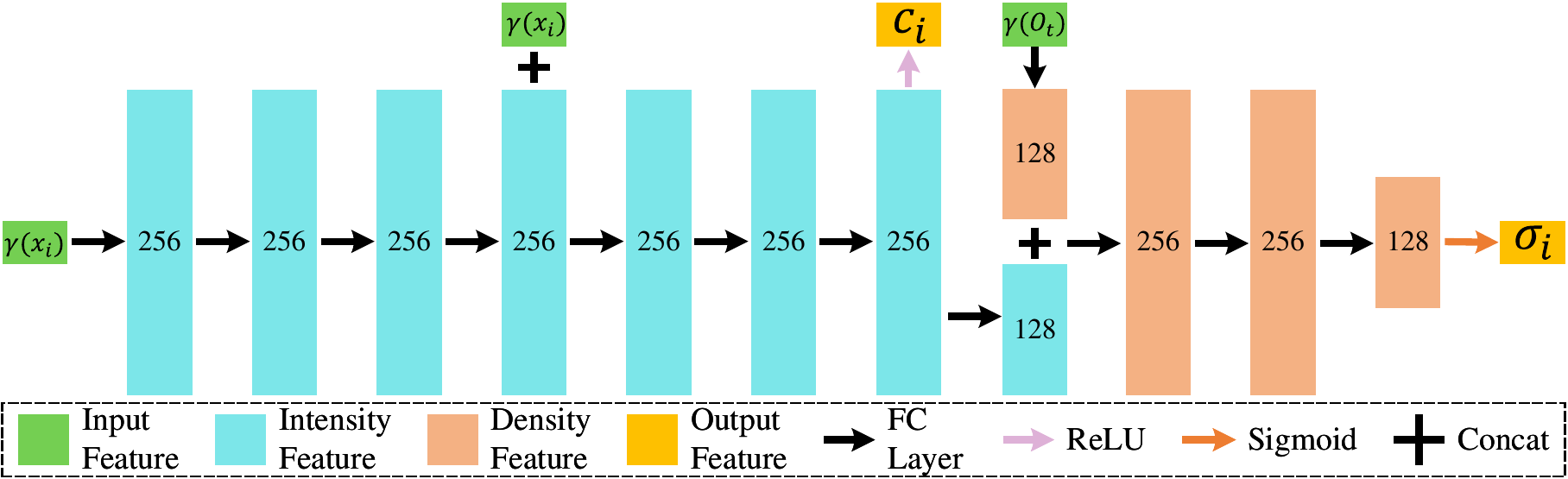}
    \caption{
    The architecture of multi-layer perceptron (MLP) $F_{\Theta}$.
    Given a center $O_t$ and sampling point $x_i$, we first encode them by the positional encoding $\gamma(\cdot)$ in Eq. \ref{eq:PE} to obtain the input feature $\gamma(O_t)$ and $\gamma(x_i)$.
    Then, we pass $\gamma(x_i)$ through 7 fully connected (FC) layers with concatenating the 4th intermediate hidden layer and the input vector itself to predict the intensity $c_i$ using ReLU activations.
    Finally, we feed $\gamma(O_t)$ into the MLP and concatenate the density feature and the 8th layer of intensity feature to obtain the output density $\sigma_t$ with Sigmoid activation.
    }
    \label{fig:MLP}
\end{figure*}

\subsection{Line-Segment Sampling}

Implicit neural representation (INR) techniques aim to learn the continuous representation of projection views from sparse-view (SV) sinograms.
However, the above analysis demonstrates that existing INR-based SVCT methods suffer from aliasing errors in the projection domains, leading to blurry results and severe artifacts.

To address this issue, we propose a novel sampling strategy: \textit{line-segment sampling}, which samples a line-segment region instead of a single point to build the correlation between adjacent projection views.
Specifically, to predict the sinogram pixel at the location $O_t$, we construct a line-segment region $\ell(O_t, \rho)$ with the center of $O_t$, where $\rho$ is the length of that line segment.
To estimate the distribution of $\ell(O_t, \rho)$, we first samples $N$ points within that sampled region as a point set $\{x_i\}_{i=1}^N$.
Each point is sampled by:
\begin{equation}
    x_i\sim U\left[\ell(O_t, \rho)\right],
\end{equation}
where $U$ denotes the uniform distribution.
Then we feed the point set $\{x_i\}_{i=1}^N$ and the center $O_t$ into a multi-layer perceptron (MLP) network $F_\Theta$ to predict a set of density $\{\sigma_i\}_{i=1}^N$ and intensity $\{c_i\}_{i=1}^N$ within the sampled region $\ell(O_t, \rho)$ by:
\begin{equation}
    \sigma_i, c_i = F_\Theta(\gamma(O_t), \gamma(x_i)),
\end{equation}
where $\gamma(\cdot)$ denotes the positional encoding introduced in \cite{NeRF}, which can be formulated as:
\begin{equation}
    \gamma(y)=y\bigcup^{\omega-1}_{i=0}(\sin(2^{i}y), \cos(2^{i}y)),\ where\ \omega\in\mathbb{N},
    \label{eq:PE}
\end{equation}
where $y$ is an arbitrary input vector, $\omega$ is set to 10 as default.
Finally, since $\sigma_i$ and $c_i$ denote the density and intensity of $x_i$, we can build the continuous function $\sigma(O_t, \cdot)$ and $c(\cdot)$ to estimate the distribution of the line-segment regions $\ell(O_t, \rho)$ via these discrete samples.
\textit{Note} the density function is related to $(O_t, x_i)$ while the intensity function is only related to $x_i$, because we assume the density distribution of each line-segment region is independent, while the intensity of all points is fixed.
% As a result, after densely sampling the points within the line-segment region $\ell(O_t, \rho)$ for optimization, the model can build the continuous representation between adjacent projections, alleviating the aliasing errors in the projection domains.

\subsection{Center-based Line Integral}

To obtain the sinogram pixel $\mathbf{C}(O_t,\rho)$ at $O_t$, an intuitive solution is to calculate the line integral as:
% \begin{equation}
%     \mathbf{C}(O_t,\rho) = \int_{-\frac{\rho}{2}}^{\frac{\rho}{2}}\sigma(O_t, \nu)c(O_t + \nu)d\nu,
%     \label{eq:line}
% \end{equation}
\begin{equation}
    \mathbf{C}(O_t,\rho) =\frac{1}{\rho\sum\sigma_i}\int_{0}^{\rho}\sigma(O_t, \nu)c(O_t + \nu)d\nu,
    \label{eq:line}
\end{equation}
where $\nu$ is the distance between each sampling point $x_i$ to the starting point $O_t-\frac{\rho}{2}$, and $\frac{1}{\rho\sum\sigma_i}$ is the normalization term to ensure the integral output within a reasonable range.
However, Eq. \ref{eq:line} neglects the image continuity priors within the sampling region $\ell(O_t,\rho)$, leading to sub-optimal results.

Inspired by \cite{CRF} that assigns the nearby pixels with similar potentials, we assume the density of each point $x_i$ within the line-segment region $\ell(O_t,\rho)$ is attenuated from the center $O_t$ to the ends.
We employ the Lambert-Beer Law to estimate the distribution of attenuation coefficients within the line-segment region $\ell(O_t, \rho)$, and thus the proposed \textit{center-based line integral} can be formulated as:
\begin{equation}
    \mathbf{C}(O_t,\rho) = \int_{0}^{\frac{\rho}{2}}\frac{c(O_t + \nu)(1-\exp(-\sigma(O_t,\nu)))}{\exp(\int_{0}^{\nu}\sigma(O_t,\nu'))d\nu'}d\nu,
\end{equation}
where $\nu=\|O_t-x_i\|$ denotes the distance between each sampling point $x_i$ and the center $O_t$.
Given $N$ sampling points by the proposed \textit{line-segment sampling}, we first sort these points by the distance between the center $O_t$ and themselves.
Subsequently, the above \textit{center-based line integral} can be approximated via numerical quadrature rules:
\begin{equation}
  \hat{\mathbf{C}}(O_t,\rho)=\sum^{N-1}_{i=1}\frac{1-\exp(-\sigma(O_t,\nu_i)(\nu_{i+1}-\nu_{i}))}{\exp(\sum_{j=1}^{i}\sigma(O_t, \nu_j)(\nu_{j+1}-\nu_{j}))}c(O_t + \nu_i),
  \label{eq:line_integral}
\end{equation}
where $\hat{\mathbf{C}}(O_t,\rho)$ denotes the approximated results of $\mathbf{C}(O_t,\rho)$.

Given the predicted result $\hat{\mathbf{C}}(O_t,\rho)$ at Eq. \ref{eq:line_integral} using $N$ sorted sampling points, \textit{APRF} can be optimized in the following MSE loss $\mathcal{L}$:
\begin{equation}
    \mathcal{L} = \frac{1}{len(\mathcal{B})}\sum_{O_t\in\mathcal{B}}\|g.t.-\hat{\mathbf{C}}(O_t,\rho)\|_2^2,
    \label{eq:loss}
\end{equation}
where $\mathcal{B}$ denotes the SV coordinates in a batch, and $g.t.$ is the corresponding SV sinogram pixels.

\begin{table*}[!t]
    \setlength{\tabcolsep}{0.6mm}
    \caption{
    Comprehensive comparisons on COVID-19 \cite{covid19} for \bt{2D} parallel- and fan-beam SVCT of the various number of projection views.
    \bt{Bold} and \underline{underline} texts indicate the best and second best performance.
    }
    \label{tab:2DSVCT}
    \centering
    \scriptsize
    \begin{tabular}{c|ccc|ccc|ccc|ccc|ccc|ccc}
    \hline
    \multirow{3}{*}{Methods} & \multicolumn{9}{c|}{\textit{Parallel}-beam SVCT Reconstruction} & \multicolumn{9}{c}{\textit{Fan}-beam SVCT Reconstruction} \\\cline{2-19}
    & \multicolumn{3}{c|}{45 \textit{views}} & \multicolumn{3}{c|}{60 \textit{views}} & \multicolumn{3}{c|}{90 \textit{views}} & \multicolumn{3}{c|}{45 \textit{views}} & \multicolumn{3}{c|}{60 \textit{views}} & \multicolumn{3}{c}{90 \textit{views}} \\\cline{2-19}
                        &PSNR\ua &SSIM\ua &LPIPS\da &PSNR\ua &SSIM\ua &LPIPS\da &PSNR\ua &SSIM\ua &LPIPS\da &PSNR\ua &SSIM\ua &LPIPS\da &PSNR\ua &SSIM\ua &LPIPS\da &PSNR\ua &SSIM\ua &LPIPS\da \\\hline
FBP \cite{FBP}          & 24.47 & 0.4043 & 0.3994  & 25.62  & 0.4794 & 0.3400  & 26.99  & 0.5773 & 0.2554  & 26.18  & 0.4043 & 0.4733  & 26.87  & 0.4805 & 0.4240  & 28.94 & 0.5969  & 0.3501   \\
FBPConvNet \cite{FBPConvNet} & 31.18 & 0.7510 & \underline{0.2111} & 31.76             & 0.7606             & \underline{0.1720} & 32.34             & 0.7659             & 0.1511             & 29.86             & 0.6550             & \underline{0.2944} & 30.83             & 0.6800             & \underline{0.2703} & 31.67             & 0.7207             & \underline{0.2390}   \\
GRFF \cite{GRFF}             & 30.64             & 0.7321             & 0.3792             & 30.40             & 0.7163             & 0.3850             & 31.35             & 0.7299             & 0.3675             & 30.08             & 0.6996             & 0.4000             & 30.53             & 0.7246             & 0.3808             & 30.68             & 0.7333             & 0.3928   \\
NeRP \cite{NeRP}             & \underline{31.48} & \underline{0.7621} & 0.3671             & 31.45             & 0.7666             & 0.3655             & 31.91             & 0.7741             & 0.3521             & \bt{31.02}    & 0.7505             & 0.3844             & \underline{31.31} & 0.7690             & 0.3770             & 31.59             & 0.7857             & 0.3637   \\
CoIL \cite{Coil}             & 30.92             & 0.7556             & 0.2677             & \underline{31.96} & \underline{0.7863} & 0.2277             & 33.28             & 0.8164             & 0.1829             & 30.79             & \bt{0.7577}    & 0.3843             & 30.66             & \underline{0.7860} & 0.3471             & \underline{31.72} & \underline{0.8186} & 0.2990   \\
SCOPE \cite{scope}           & 29.87             & 0.7072             & 0.2723             & 31.58             & 0.7570             & 0.2071             & \underline{33.64} & \underline{0.8181} & \underline{0.1219} & 24.81             & 0.6385             & 0.3729             & 25.00             & 0.6577             & 0.3749             & 25.39             & 0.6780             & 0.3412   \\
\bt{APRF} (Ours)       &\bt{31.92} &\bt{0.7655} &\bt{0.2084} &\bt{33.43} &\bt{0.8009} &\bt{0.1665} &\bt{34.74} &\bt{0.8342} &\bt{0.1131} &\underline{30.42} &\underline{0.7545} &\bt{0.2871} &\bt{31.59} &\bt{0.7948} &\bt{0.2311} &\bt{32.48}    & \bt{0.8305}    & \bt{0.2116}  \\\hline
    \end{tabular}
\end{table*}

\subsection{Spatial Normalization \& Multi-Layer Perceptron}
\noindent\bt{Spatial Normalization.}
To build the coordinate-based representations field for the given SV sinograms, we first normalize the coordinates from the projection domains to the field spaces within the range of $(-1, 1)$ to adapt the positional encoding $\gamma(\cdot)$ in Eq. \ref{eq:PE}.
For example, given a $L$$\times$$H$$\times$$W$ 3D sinogram, we convert the sinogram coordinate $(l, h, w)$ into the corresponding field coordinate $O_t$ by:
\begin{equation}
    O_t=(\frac{2l-L}{L+2P}, \frac{2h-H}{H+2P}, \frac{2w-W}{W+2P}, )
\end{equation}
where $P$ set to $1$ is a padding size to restrict $O_t$ within the range of $(-1, 1)$, $L$ denotes the number of projection views, $H$ and $W$ indicate the height and width of detector plane.
\noindent\bt{The Architecture of Multi-Layer Perceptron.}
We reform the multi-layer perceptron (MLP) architecture introduced in \cite{NeRF} to adapt the proposed \textit{line-segment sampling}.
As shown in Figure \ref{fig:MLP}, we employ the positional encoding in Eq. \ref{eq:PE} to enrich the feature of $O_t$ and $x_i$ before feeding them into the MLP.
Since $\sigma_i$ is related to $(O_t, x_i)$ while $c_i$ is only corresponding to $x_i$, we decouple the features of $\sigma_i$ and $c_i$.
Specifically, we first feed $\gamma(x_i)$ into the MLP to predict the intensity $c_i$, and then we concatenate the features of intensity and the input center $\gamma(O_t)$ to predict the density $\sigma_i$.
Because the proposed \textit{center-based line integral} is differentiable, the MLP can be optimized by minimizing the MSE loss $\mathcal{L}$ in Eq. \ref{eq:loss} between the predicted results and SV sinogram pixels.

\section{Experiments}

In this section, we conduct extensive experiments and in-depth analysis to demonstrate the superiority of the proposed \textit{APRF} for sparse-view computed tomography (SVCT) reconstruction.
We utilize the same hyperparameters and model settings in all experiments for fair comparisons.

\subsection{Experimental Details}

\noindent\bt{Compared Methods.}
We conduct a comprehensive comparison over 6 state-of-the-art methods, including 1 conventional analytical reconstruction method: FBP \cite{FBP}; 1 fully-supervised deep learning-based method: FBPConvNet \cite{FBPConvNet}; and 4 recent self-supervised INR-based methods: CoIL \cite{Coil}, SCOPE \cite{scope}, GRFF \cite{GRFF} and NeRP \cite{NeRP}.
FBP is re-implemented by ODL \cite{odl}, we follow the settings released on the official tutorials\footnote{https://github.com/odlgroup/odl/tree/master/examples/tomo} to deal with the re-projection of \textit{parallel}, \textit{fan} and \textit{cone} X-ray beam sinograms.
For the fully-supervised methods: FBPConvNet\footnote{https://github.com/panakino/FBPConvNet} \cite{FBPConvNet}, we use the MatConvNet toolbox \cite{matlab} to train them on the training set extracted from COVID-19 \cite{covid19} dataset.
Specifically, for each X-ray beam (\textit{parallel}, \textit{fan} and \textit{cone} X-ray beams) and projection number (45, 60, and 90), we re-train a specific model for the corresponding CT reconstruction.
The maximum iteration is set to $100$ (epochs) and the learning rate is decreasing logarithmically from $1e^{-2}$ to $1e^{-3}$.
We employ Adam \cite{adam} as the optimizer and batch size is set to $8$.
For the self-supervised INR-based methods, we directly train them on a single SV sinogram itself to reconstruct CT images.
Since the output of CoIL \cite{Coil}, SCOPE \cite{scope} and our \textit{APRF} is a DV sinogram, we apply the same re-projection process on the predicted DV sinograms to obtain the CT images.
\textit{Note} that the results of CoIL\footnote{https://github.com/wustl-cig/Cooridnate-based-Internal-Learning} \cite{Coil}, SCOPE\footnote{https://github.com/iwuqing/SCOPE} \cite{scope}, GRFF\footnote{https://github.com/tancik/fourier-feature-networks} \cite{GRFF} and NeRP\footnote{https://github.com/liyues/NeRP} \cite{NeRP} are obtained using their official implementations.

\begin{figure*}[!t]
    \centering
    \includegraphics[width=\textwidth]{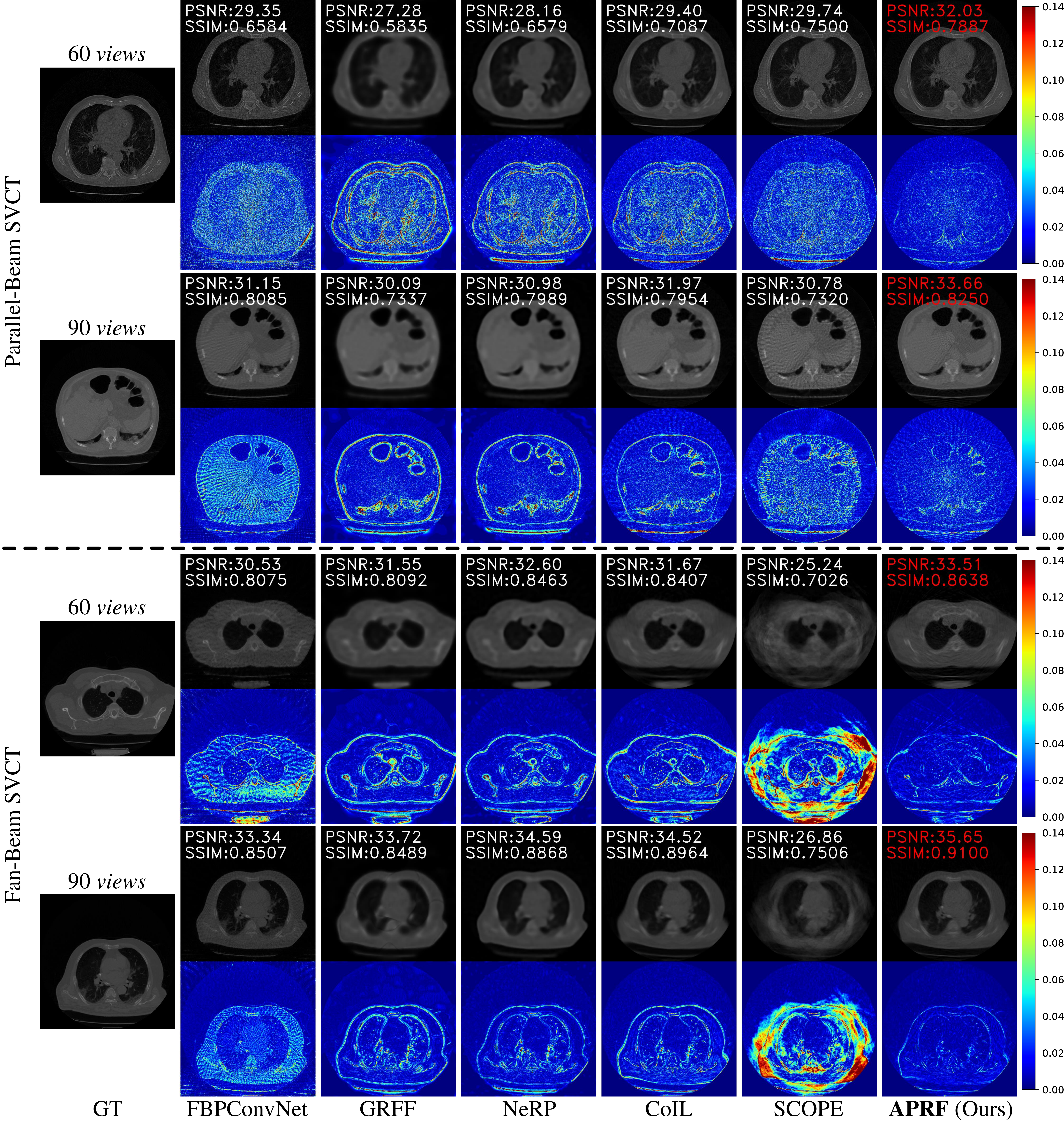}
    \caption{
    Visual comparisons between our \textit{APRF} and 5 state-of-the-art methods: FBPConvNet \cite{FBPConvNet}, GRFF \cite{GRFF}, NeRP \cite{NeRP}, CoIL \cite{Coil} and SCOPE \cite{scope} for \textbf{2D} SVCT reconstruction on COVID-19 \cite{covid19} dataset.
    Heatmaps at the bottom visualize the difference related to the GT images.
    }
    \label{fig:visual_2d}
    % \vspace{-2em}
\end{figure*}

\noindent\bt{Datasets. } 
The experimental data are selected from 2 publicly available CT datasets: COVID-19 \cite{covid19} and 2019 Kidney Tumor Segmentation Challenge (KiTS19) \cite{kits19} datasets.
Since the proposed \textit{APRF} is the self-supervised method, \emph{i.e.,} only trained on the test CT image itself at test time without the demands of large training data, we do not require to build an extra dataset for training.

\bt{1)} \bt{COVID-19} \cite{covid19} dataset, containing considerable 3D CT volumes from 1000+ subjects with confirmed COVID-19 infections, is a large-scale CT dataset for the clinical study of COVID-19 virus.
We randomly select 600 2D slices from 100 subjects, each slice is extracted on axial view from the 3D CT volumes.
Specifically, we select 500 slices from 50 subjects as the training set, 60 2D slices from 10 subjects as the validation set, and 40 2D slices from the rest 40 subjects as the test set.
All the 2D CT slices have the same image dimension of 512$\times$512.
\textit{Note} that the training and validation sets are only employed to train the fully-supervised methods: FBPConvNet \cite{FBPConvNet}, while FBP \cite{FBP}, CoIL \cite{Coil}, SCOPE \cite{scope}, GRFF \cite{GRFF}, NeRP \cite{NeRP} and the proposed \textit{APRF}, are only trained on the test SV sinograms itself to reconstruct CT images.

\bt{2)} \bt{2019 Kidney Tumor Segmentation Challenge} (KiTS19) \cite{kits19} dataset consists of arterial phase abdominal CT scans from 300 unique kidney cancer subjects who underwent partial or radical nephrectomy.
\textit{Note} that FBPConvNet \cite{FBPConvNet} is only designed for 2D SVCT reconstruction, and thus we only prepare the test set to evaluate the performance of FBP \cite{FBP} and other self-supervised INR-based methods.
Specifically, we randomly select 40 3D CT volumes, each 3D CT volume is resized into 256$\times$256$\times$80 image dimension.

\noindent\bt{Dataset Simulation.}
For thoroughly evaluating the reconstruction quality of SVCT with various X-ray beams, we follow the strategies in \cite{Coil,NeRP} to simulate \textit{parallel}, \textit{fan} and \textit{cone} X-ray beam sparse-view (SV) sinograms by projecting the raw CT images using Operator Discretization Library (ODL) \cite{odl}.
Specifically, we follow the above-mentioned official settings to extract \textit{parallel-} and \textit{fan-}beam SV sinograms from 2D CT slices, and generate \textit{parallel-} and \textit{cone-}beam SV sinograms from 3D CT volumes.
Each SV sinogram is generated at different projection views (45, 60, and 90).
Following the self-supervised settings, \textit{APRF} is trained on a test SV sinogram itself, while the raw CT images are only used as ground truths for evaluation.
\textit{Note} that the \textit{parallel}, \textit{fan} and \textit{cone} X-ray beam SVCT are considered as three independent reconstruction tasks, and thus we solely conduct the training and test processes of each X-ray beam.

\begin{table*}[!t]
    \setlength{\tabcolsep}{0.6mm}
    \caption{
    Comprehensive comparisons on KiTS19 \cite{kits19} for \bt{3D} parallel- and cone-beam SVCT of the various number of projection views.
    \bt{Bold} and \underline{underline} texts indicate the best and second best performance.
    }
    \label{tab:3DSVCT}
    \centering
    \scriptsize
    \begin{tabular}{c|ccc|ccc|ccc|ccc|ccc|ccc}
    \hline
    \multirow{3}{*}{Methods} & \multicolumn{9}{c|}{\textit{Parallel}-beam SVCT Reconstruction} & \multicolumn{9}{c}{\textit{Cone}-beam SVCT Reconstruction} \\\cline{2-19}
    & \multicolumn{3}{c|}{45 \textit{views}} & \multicolumn{3}{c|}{60 \textit{views}} & \multicolumn{3}{c|}{90 \textit{views}} & \multicolumn{3}{c|}{45 \textit{views}} & \multicolumn{3}{c|}{60 \textit{views}} & \multicolumn{3}{c}{90 \textit{views}} \\\cline{2-19}
    & PSNR\ua    & SSIM\ua     & LPIPS\da  & PSNR\ua    & SSIM\ua     & LPIPS\da  & PSNR\ua    & SSIM\ua     & LPIPS\da  & PSNR\ua    & SSIM\ua     & LPIPS\da  & PSNR\ua    & SSIM\ua     & LPIPS\da  & PSNR\ua    & SSIM\ua     & LPIPS\da    \\\hline
    % \multicolumn{19}{c}{Conventional Methods} \\\hline 
    FBP \cite{FBP}        & 24.74             & 0.6324             & 0.2856             & 25.49             & 0.6923             & 0.2114             & 26.21             & 0.7499             & \underline{0.1124} & 24.74             & 0.6324             & \underline{0.2856} & 25.45             & 0.6923             & \underline{0.2114} & 26.21             & 0.7499             & \underline{0.1724}   \\
    % \multicolumn{19}{c}{Self-Supervised Methods} \\\hline            
    GRFF \cite{GRFF}      & 29.83             & \underline{0.8455} & 0.1922             & 29.74             & 0.8300             & 0.1873             & 29.97             & 0.8349             & 0.1845             & 27.75             & 0.7296             & 0.3605             & 28.76             & 0.7382             & 0.3271             & 28.79             & 0.7541             & 0.3387   \\
    NeRP \cite{NeRP}      & \underline{30.13} & 0.8436             & \underline{0.1789} & \underline{30.89} & \underline{0.8587} & \underline{0.1685} & \underline{31.04} & \underline{0.8601} & 0.1627             & \underline{29.03} & \underline{0.7854} & 0.2878             & \underline{29.62} & \underline{0.7938} & 0.2744             & 29.94             & \underline{0.8003} & 0.2652   \\
    CoIL \cite{Coil}      & 28.86             & 0.7466             & 0.3015             & 29.15             & 0.7573             & 0.2881             & 29.72             & 0.7807             & 0.2663             & 27.76             & 0.7378             & 0.4047             & 28.16             & 0.7510             & 0.3904             & \underline{30.42} & 0.7807             & 0.2663   \\
    \bt{APRF} (Ours) & \bt{32.51}    & \bt{0.8633}    & \bt{0.1138}    & \bt{33.89}    & \bt{0.8982}    & \bt{0.0782}    & \bt{34.12}    & \bt{0.9060}    & \bt{0.0739}    & \bt{29.51}    & \bt{0.7939}    & \bt{0.2470}    & \bt{31.16}    & \bt{0.8618}    & \bt{0.1792}    & \bt{32.08}    & \bt{0.8703}    & \bt{0.1142}   \\\hline
    \end{tabular}
  \end{table*}

\begin{figure*}[!t]
    \centering
    \includegraphics[width=\textwidth]{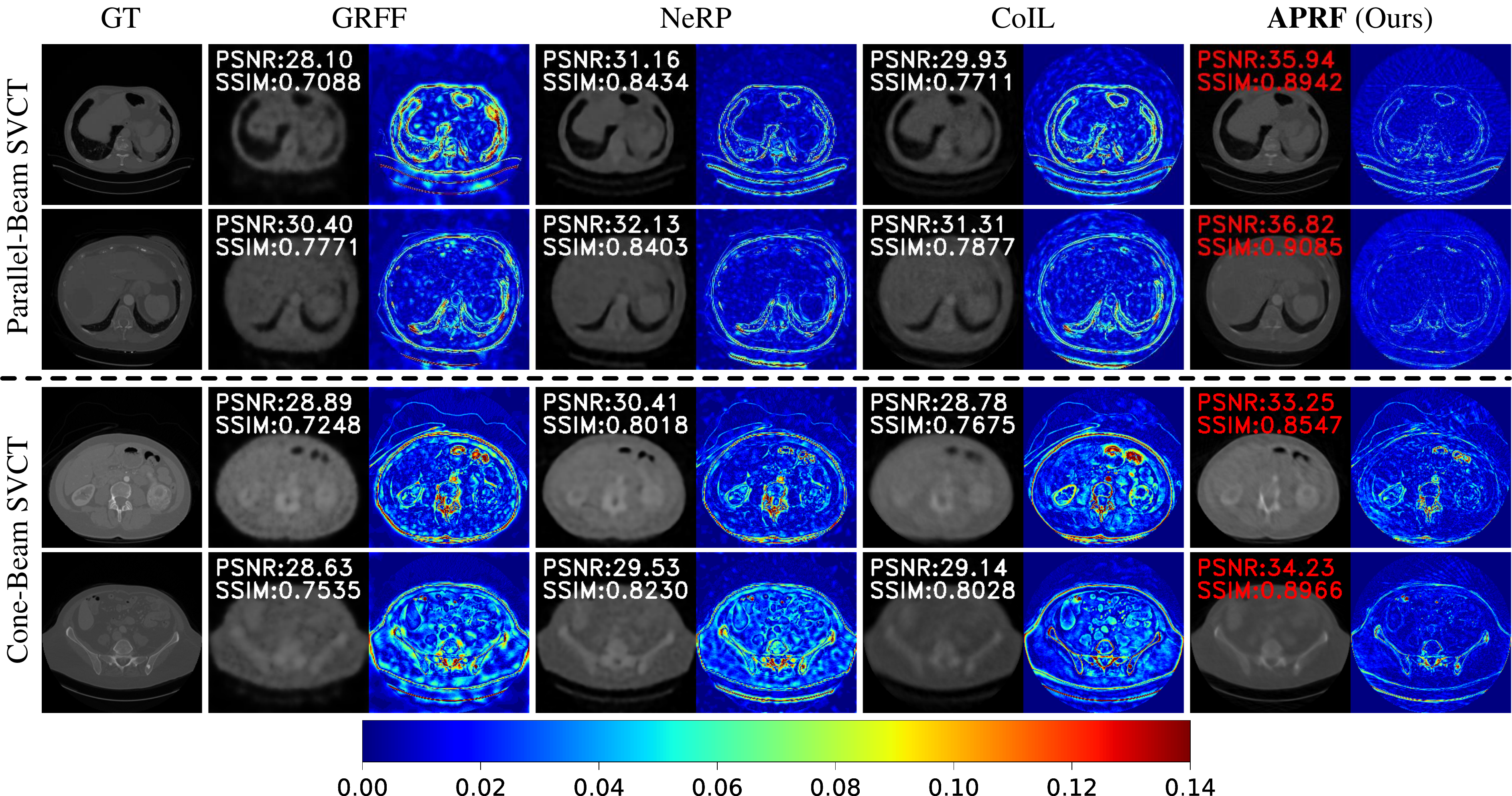}
    \caption{
    Visual comparisons between our \textit{APRF} and 3 state-of-the-art methods: GRFF \cite{GRFF}, NeRP \cite{NeRP} and CoIL \cite{Coil} for \textbf{3D} SVCT reconstruction on KiTS19 \cite{covid19} dataset.
    Top and down rows in each type of SVCT are the results of 60 and 90 \textit{views}, respectively.
    Heatmaps at the right side visualize the difference related to the GT images.
    }
    \label{fig:visual_3d}
    % \vspace{-2em}
\end{figure*}

\noindent\bt{Evaluation Metrics. }
We employ three commonly-used objective image quality metrics: Peak Signal-to-Noise Ratio (PSNR), Structural Similarity Index Measure (SSIM) \cite{ssim}, and LPIPS \cite{lpips} to evaluate the performance of the compared methods for 2D and 3D SVCT reconstruction.
PSNR is based on pixel-to-pixel distance while SSIM utilizes the mean and variance of images to measure structural similarity.
Unlike the above conventional metrics, LPIPS is an objective perceptual similarity metric based on deep learning.
In our experiments, PSNR and SSIM are re-implemented by scikit-image \cite{skimage} packages, while we use the default settings of LPIPS for evaluations.

\noindent\bt{Implementation Details.} \label{sec:details}
Our method is implemented on top of \cite{nerf-pytorch}, a Pytorch \cite{Pytorch} re-implementation of NeRF \cite{NeRF}.
Meanwhile, our experiments are also based on Pytorch \cite{Pytorch} framework, and run on a single NVIDIA RTX A6000 GPU.
The length $\rho$ of the line segment $\ell(O_t,\rho)$ is set to $\frac{2}{L+P}$, which means it depends on the number of projection views (\emph{i.e.,} long in SV sinograms but short in DV ones).

\textit{For training}, we sample $N_{train}=65$ points within a line segment $\ell(O_t,\rho)$ and then feed the sampling points into the multi-layer perceptron (MLP) $F_{\Theta}$.
We employ Adam \cite{adam} as the optimizer with a weight decay $1e^{-7}$ and a batch size $2048$.
The maximum iteration is $40000$ for all CT images and the learning rate is annealed logarithmically from $3e^{-3}$ to $2e^{-5}$.

\textit{For test}, we first generate DV sinograms by uniformly sampling 720 partitions within $[0, \pi]$, and then apply FBP \cite{FBP} re-implemented by ODL \cite{odl} to reconstruct the CT images.
The number of sampling points is set to $N_{test}=\frac{(N_{train} - 1)\eta}{720} + 1$, where $\eta$ is the number of projection views preseted in training.
It means the number of test sampling points is 16$sim$8 times smaller than in training for the 45$\sim$90-view sinograms, reducing considerable computational costs.

\noindent\bt{Runtime \& Parameters.}
The average training time of the proposed \textit{APRF} for 2D SVCT reconstruction is about 5 \textit{mins} while training for reconstructing 3D SVCT demands about 15 \textit{mins}.
Since the number of sampling points in the test stage is much smaller than in training, the inference time for a single 2D SV sinogram with 800$\times$512 size is only about 1.4 \textit{secs}, while synthesizing a 3D SV sinogram with 400$\times$400$\times$80 size requires about 40 \textit{secs}.
It is worth noting that \textit{APRF} is a lightweight model, which only contains 0.55M parameters.

\begin{table*}[!t]
    \caption{Comprehensive comparisons of various ablation variants for various X-ray beam SVCT reconstruction of \textit{60} projection views.
  \bt{Bold} text indicates the best performance in this table.
  }
    \label{tab:ab}
    \centering
    % \normalsize
    \begin{tabular}{cccc|ccc|ccc}
    \hline
    \multirow{3}{*}{\textit{LSS}} & \multirow{3}{*}{\textit{CLI}} & \multirow{3}{*}{\textit{cnt.}} & \multirow{3}{*}{\textit{dec.}} & \multicolumn{6}{c}{2D SVCT Reconstruction}                                                         \\\cline{5-10}
    & & & & \multicolumn{3}{c|}{\textit{Parallel} X-ray beam} & \multicolumn{3}{c}{\textit{Fan} X-ray beam} \\\cline{5-10}
        &    &    &    & PSNR\ua              & SSIM\ua                & LPIPS\da               & PSNR\ua              & SSIM\ua                & LPIPS\da               \\\hline
        &    &    &    & 30.53                & 0.7488                 & 0.2517                 & 29.15                & 0.7266                 & 0.3105                 \\
    \ck &    &    &    & 31.59\df{+1.06}      & 0.7642\df{+.0154}      & 0.2019\df{-.0498}      & 30.14\df{+0.99}      & 0.7528\df{+.0262}      & 0.2708\df{-.0397}      \\
    \ck &\ck &    &    & 32.62\df{+1.03}      & 0.7884\df{+.0242}      & 0.1776\df{-.0243}      & 30.98\df{+0.84}      & 0.7798\df{+.0270}      & 0.2445\df{-.0263}      \\
    \ck &\ck &\ck &    & 33.05\df{+0.43}      & 0.7948\df{+.0064}      & 0.1688\df{-.0074}      & 31.31\df{+0.33}      & 0.7884\df{+.0086}      & 0.2391\df{-.0054}      \\
    \ck &\ck &\ck &\ck & \bt{33.43}\df{+0.38} & \bt{0.8009}\df{+.0061} & \bt{0.1665}\df{-.0037} & \bt{31.59}\df{+0.28} & \bt{0.7948}\df{+.0064} & \bt{0.2311}\df{-.0080} \\\hline
    \multirow{3}{*}{\textit{LSS}} & \multirow{3}{*}{\textit{CLI}} & \multirow{3}{*}{\textit{cnt.}} & \multirow{3}{*}{\textit{dec.}} & \multicolumn{6}{c}{3D SVCT Reconstruction} \\\cline{5-10}
    & & & & \multicolumn{3}{c|}{\textit{Parallel} X-ray beam} & \multicolumn{3}{c}{\textit{Cone} X-ray beam} \\\cline{5-10}

        &    &    &    & PSNR\ua              & SSIM\ua                & LPIPS\da               & PSNR\ua              & SSIM\ua                & LPIPS\da  \\\hline
        &    &    &    & 30.22                & 0.8183                 & 0.1756                 & 27.97                & 0.7663                 & 0.3286 \\
    \ck &    &    &    & 31.37\df{+1.15}      & 0.8435\df{+.0252}      & 0.1372\df{-.0384}      & 29.23\df{+1.26}      & 0.8153\df{+.0490}      & 0.2609\df{-.0677} \\
    \ck &\ck &    &    & 32.69\df{+1.32}      & 0.8797\df{+.0362}      & 0.0984\df{-.0388}      & 30.17\df{+0.94}      & 0.8428\df{+.0275}      & 0.2157\df{-.0452}  \\
    \ck &\ck &\ck &    & 33.25\df{+0.56}      & 0.8940\df{+.0093}      & 0.0892\df{-.0092}      & 30.78\df{+0.61}      & 0.8515\df{+.0087}      & 0.1949\df{-.0208}  \\
    \ck &\ck &\ck &\ck & \bt{33.89}\df{+0.64} & \bt{0.8982}\df{+.0092} & \bt{0.0782}\df{-.0110} & \bt{31.16}\df{+0.38} & \bt{0.8618}\df{+.0103} & \bt{0.1792}\df{-.0157}  \\\hline
    \end{tabular}
\end{table*}

\begin{table*}[!t]
    \caption{Comprehensive comparisons under different parametric settings for various X-ray beam SVCT reconstruction of \textit{60} projection views.
    \bt{Bold} and \underline{underline} texts indicate the best two results in this table.
  }
    \label{tab:ps}
    \centering
    \footnotesize
    \begin{tabular}{c|ccc|ccc|ccc|ccc}
    \hline
    \multirow{3}{*}{} & \multicolumn{6}{c|}{2D SVCT Reconstruction}                            & \multicolumn{6}{c}{3D SVCT Reconstruction}                             \\\cline{2-13}
    & \multicolumn{3}{c|}{\textit{Parallel} X-ray beam} & \multicolumn{3}{c|}{\textit{Fan} X-ray beam} & \multicolumn{3}{c|}{\textit{Parallel} X-ray beam} & \multicolumn{3}{c}{\textit{Cone} X-ray beam} \\\cline{2-13}
    & PSNR\ua    & SSIM\ua     & LPIPS\da  & PSNR\ua    & SSIM\ua     & LPIPS\da  & PSNR\ua    & SSIM\ua     & LPIPS\da  & PSNR\ua    & SSIM\ua     & LPIPS\da   \\\hline
    \textit{default}         & \underline{33.43} & \underline{0.8009} & \underline{0.1665} & \underline{31.59} & \bt{0.7948} & \bt{0.2311} & \underline{33.89} & \underline{0.8982} & \underline{0.0782} & \underline{31.16} & \bt{0.8618} & \underline{0.1792} \\
    $\rho\!=\!\frac{1}{L+P}$ & 32.80 & 0.7921 & 0.1785 & 30.62 & 0.7785 & 0.2447 & 31.99 & 0.8889 & 0.0912 & 30.64 & 0.8577 & 0.1875        \\
    $\rho\!=\!\frac{4}{L+P}$ & 33.22 & 0.7969 & 0.1684 & 31.27 & 0.7892 & 0.2378 & 33.45 & 0.8933 & 0.0841 & 30.47 & 0.8540 & 0.1929        \\
    $N_{test}\!=\!N_{train}$ & \bt{33.68} & \bt{0.8043} & \bt{0.1627} & \bt{31.84} & \underline{0.7941} & \underline{0.2324} & \bt{34.04} & \bt{0.9008} & \bt{0.0759} & \bt{31.31} & \underline{0.8601} & \bt{0.1774}  \\\hline
    \end{tabular}
\end{table*}

\subsection{Comparison with State-of-the-art Methods} \label{sec:comp}
In this subsection, we first report the quantitative and visual comparisons between the proposed \textit{APRF} and the 6 above-mentioned state-of-the-art methods.
Subsequently, we summarise and analyze the experimental results.

\noindent\bt{Quantitative Comparison.}
As reported in Tables \ref{tab:2DSVCT} and \ref{tab:3DSVCT}, the proposed \textit{APRF} favorably surpasses all the competitors for 2D and 3D SVCT reconstruction, yielding consistent preferable performance under the various number of projection views.
% For the reconstruction of \textit{fan}- and \textit{cone}-beam SVCT, the PSNR score of \textit{APRF} is slightly inferior to NeRP \cite{NeRP} on 2D \textit{fan}-beam SVCT of 45 \textit{views}, but outperform all the exhibited methods under the rest settings.
The outperformance suggests \textit{APRF} achieves better sample efficiency than state-of-the-art methods, obtaining higher reconstruction quality under the same number of projection views.
Especially, compared to the exhibited methods, the performance of \textit{APRF} has significant growth with the increasing number of projection views.
Moreover, the experimental results also demonstrate that \textit{APRF} can handle various X-ray beams, outperforming the other methods for \textit{parallel-}, \textit{fan-} and \textit{cone}-beam SVCT reconstruction.

\noindent\bt{Visual Comparison.}
We visualize the results of \textit{APRF} and other competitors for 2D and 3D SVCT reconstruction in Figures \ref{fig:visual_2d} and \ref{fig:visual_3d}, respectively.
As shown, the difference maps of GRFF \cite{GRFF} and NeRP \cite{NeRP} contain conspicuous artifacts at object boundaries, suggesting GRFF \cite{GRFF} and NeRP \cite{NeRP} generate blurry contents in their results.
In contrast, though CoIL \cite{Coil} and SCOPE \cite{scope} preserve more details, they also produce severe artifacts in the inner regions of objects.
Compared to the exhibited methods, the proposed \textit{APRF} achieves better visual verisimilitude to the GT images, producing more accurate details and fewer artifacts.

\subsection{Ablation Study \& Parametric Sensitivity Analysis} \label{sec:ab}
In this subsection, we conduct comprehensive experiments to prove the correctness of the model design.
We first carry out an ablation study to investigate the effectiveness of the proposed modules.
Subsequently, we evaluate the \textit{APRF}'s parametric sensitivity under different parameter settings.
For fair comparisons, each variant is trained with the same experimental settings, including the same positional encoding in Eq. \ref{eq:PE}, loss in Eq. \ref{eq:loss}, and the maximum iterations.

\noindent\bt{APRF's Ablation Variants.}
We conduct a thorough evaluation against ablation variants of the proposed \textit{APRF} with each module: \textit{LSS}, \textit{CLI}, \textit{cnt.} and \textit{dec.}, which indicate the \textit{line-segment sampling}, \textit{center-based line integral}, adding the center $O_t$ into input vector, and decoupling the density and intensity features in MLP, respectively.
The baseline model is directly feeding the points to synthesize the corresponding sinogram pixels.
For fair comparisons, the MLP of each ablation variant has a similar parameter size ($\pm$0.02M).
As reported in Table \ref{tab:ab}, our baseline model (row 1) can be conspicuously improved by the proposed \textit{LSS} (row 2), outperforming the most state-of-the-art methods on 60 projection views.
Employing the proposed \textit{CLI} instead of the line integral introduced in Eq. \ref{eq:line} can also significantly enhance the reconstruction performance (row 3).
Besides, since \textit{cnt.} and \textit{dec.} (rows 4 and 5) can further improve the reconstruction quality under similar computational costs, demonstrating the effectiveness of our modification against input vectors and MLP architecture.

\noindent\bt{APRF under Different Parameteric Settings.}
To analyze the parameteric sensitivity of the proposed \textit{APRF}, we evaluate its performance under different parameteric settings: ``$\rho\!=\!\frac{1}{L+P}$'', ``$\rho\!=\!\frac{4}{L+P}$'', and ``$N_{test}\!=\!N_{train}$'', where ``$\rho\!=\!\frac{1}{L+P}$'' and ``$\rho\!=\!\frac{4}{L+P}$'' denote the different length of line-segment regions $\ell(O_t, \rho)$, while ``$N_{test}\!=\!N_{train}$'' indicates using the same number of sampling points as $N_{train}$ in test stage.
The default setting is introduced in Section \ref{sec:details}, where $\rho\!=\!\frac{2}{L+P}$ and $N_{test}\!=\!\frac{(N_{train} - 1)\eta}{720}\! + \!1$.
Since the default length can only cover the whole coordinate-based representation fields, the 0.5$\times$ length: ``$\rho\!=\!\frac{1}{L+P}$'' may leave some unmodeled spaces in the fields, while the 2$\times$ length: ``$\rho\!=\!\frac{4}{L+P}$'' may produce blurry results.
Besides, the default number of $N_{test}$ is 8$\sim$16$\times$ smaller than $N_{train}$ for 45$\sim$90 projection views.
As reported in Table \ref{tab:ps}, the default version (row 1) achieves consistent preferable performance at various X-ray beams.
As expected, ``$\rho\!=\!\frac{1}{L+P}$'' and ``$\rho\!=\!\frac{4}{L+P}$'' (rows 2 and 3) are inferior to the default one, but they still achieve comparable performance to state-of-the-art methods on 60 projection views.
Though ``$N_{test}\!=\!N_{train}$'' can slightly improve the default performance (row 4), it takes 8$\sim$16 multiple of computational costs to deal with 45$\sim$90 projection views, which is not cost-efficient.

\subsection{Analysis}
We compare our \textit{APRF} with 6 state-of-the-art methods in Section \ref{sec:comp}.
The performance on quantitative and visual comparisons confirms the correctness of our motivation and demonstrates the effectiveness of the proposed \textit{APRF}.
It is also worth noting that \textit{APRF} can deal with different projection settings, including different types of X-ray beams and the various number of projection views, which means \textit{APRF} acquires high-quality CT images under various situations.
Moreover, we also thoroughly evaluate the performance of our \textit{APRF} under different ablations and parametric settings in Section \ref{sec:ab}.
The ablation results demonstrate the effectiveness of our model designs, while the consistent preferable performance under different parametric settings suggests our \textit{APRF} is a parameter-insensitive method.
Compared to state-of-the-art methods, \textit{APRF} achieves superior performance and better robustness against various situations, which has broader application scenarios in practice.

\section{Conclusion}
In this paper, we observed that existing Implicit Neural Representation-based methods suffer from aliasing issues in the projection domains, leading to blurry results and severe artifacts.
Based on our findings, we propose a novel method -- Anti-Aliasing Projection Representation Field (APRF), which aims to alleviate the aliasing errors in reconstructing CT images from sparse-view (SV) sinograms.
Specifically, \textit{APRF} first employs \textit{line-segment sampling} to estimate the distribution of projections within a line segment.
Then, \textit{APRF} synthesize the corresponding sinogram values using \textit{center-based line integral}.
After training \textit{APRF} on a single SV sinogram itself, the internal regions between adjacent projection views can be modeled by spatial constraints.
As a result, \textit{APRF} can synthesize the corresponding dense-view sinograms with consistent continuity and yield high-quality CT images via the re-projection techniques.
Comprehensive experiments on CT images demonstrate that \textit{APRF} surpasses state-of-the-art methods for sparse-view reconstruction, producing better visual verisimilitude and fewer artifacts.

\bibliographystyle{IEEEtran}
\bibliography{egbib}
\end{document}